\documentclass[12pt,
    aps,
    prb,           
    onecolumn,	
    nofootinbib    
]{revtex4-2}

\usepackage{amsmath,amssymb,amsfonts}%
\usepackage{amsthm}%
\usepackage{dsfont}
\usepackage{mathrsfs}%
\usepackage{array}[=2016-10-06]
\usepackage{tabularx}%
\usepackage{graphicx}%
\usepackage{textcase}
\usepackage{mathtools}

\usepackage[colorlinks=true, allcolors=blue]{hyperref} 

\usepackage{setspace}
\begin{document}

\title{Approximate thermodynamics of the two-dimensional Ising model in an external magnetic field}
\thanks{Published in \textit{Journal of Physics A: Mathematical and Theoretical}, DOI: 10.1088/1751-8121/adcd32. Corrected version.}

\author{Mathieu Gaudreault}
\affiliation{Sir Peter MacCallum Department of Oncology, the University of Melbourne, Victoria 3000, Australia}
\affiliation{Peter MacCallum Cancer Centre, Melbourne, Victoria 3000, Australia}
\email{mathieu.gaudreault@unimelb.edu.au}

\date{\today}

\begin{abstract}

We introduce approximate expressions for the thermodynamics of the two-dimensional Ising model in an external magnetic field. The external field breaks the system's symmetry, which complicates the exact calculation of the free energy. To address this, we write the model in its quaternion representation to apply a small angle approximation with respect to a reference frame. The approximation simplifies the transfer matrix to a centrosymmetric matrix for which the largest eigenvalue is known and depends on the external field strength. We therefore write an approximate expression for the system's free energy and derive the magnetization, susceptibility, internal energy, and specific heat. These results are compared with numerical simulations performed with the BKL algorithm. We find that the approximate expressions and numerical simulations agree in regions where the approximation is valid. 
\end{abstract}

\maketitle


\section{Introduction}

The Ising model is a fundamental system for studying phase transitions and critical phenomena. It was initially developed to model ferromagnetism in materials under the influence of an external magnetic field of strength $H$ \cite{re:ising25}. In its simplest form, the magnetic moment of each atom is modeled by a discrete variable named spin, which can have two possible states, +1 or -1. At low temperatures, spins tend to be aligned, and the phase is ordered. At high temperatures, spins fluctuate, and the phase is disordered \cite{re:chandler82}. Since its introduction, the Ising model has been solved in one dimension \cite{re:ising25} and in two dimensions with no applied field ($H = 0$) \cite{re:onsager44}. Due to its simplicity, the model extends ferromagnetism, finding applications in several physical systems, including lattice-gas-driven systems \cite{re:odor03}, spin glasses \cite{re:binder76}, and protein folding \cite{re:pande97,re:gaudreault09}. Moreover, its broad applicability has made the Ising model a valuable tool in multiple fields such as biology \cite{re:mello21,re:torquato11}, neuroscience \cite{re:chialvo10}, and economics \cite{re:sornette14}. 

The two-dimensional Ising model in an external field ($H \neq 0$) is unsolved. Consequently, no analytical expressions exist to describe the system's thermodynamics. The challenge comes from the addition of the external magnetic field. In this case, symmetry is broken and the transfer matrix is no longer centrosymmetric, which complicates the derivation of the system's free energy. Nevertheless, progress has been made with perturbation theory at small and large temperatures \cite{re:mccoy78,re:zamolodchikov11,re:meissner23} and conformal field theory with integrable perturbations \cite{re:belavin84,re:xu22}. Of particular interest is to obtain a closed expression for the susceptibility of the model to study its behavior near the critical temperature \cite{re:kong86,re:Orrick01}.
 
In the exact derivation of the free energy with $H = 0$, Onsager introduced a geometric representation of the model by using the quaternion basis \cite{re:onsager44}. This representation has also been used by Kaufman to write the system as plane rotations in a multi-dimensional space \cite{re:kaufman49a}. This representation is useful as it allows the partition function to be written as a transfer matrix.  We hypothesize that advances in the derivation of the free energy of the two-dimensional model in an external field ($H \neq 0$) can be made by using a rotated frame in the quaternion representation. By considering a small angle approximation, we propose to derive approximate expressions for the system's thermodynamics dependent on the external magnetic field strength.

We provide approximate expressions for the thermodynamics of the two-dimensional Ising model in an external magnetic field. The analytical expressions are compared with the results of numerical simulations performed with the BKL algorithm of the exact model. For completeness, the approximation is also applied to the one-dimensional Ising model in the supplementary material.

\section{Statistical mechanics of the Ising model} 

Consider the Ising model where spins can have two states $S = \pm 1$. Assume that spins can interact with their nearest neighbors with energy $J > 0$. In an external magnetic field of strength $H$, the system's energy $E$ is \cite{re:mccoy73,re:baxter82,re:huang87}
\begin{equation}
E = -J\sum_{\langle i,j \rangle} S_{i}S_{j} - H\sum_{i} S_{i} \;,
\label{eq:energy_state}
\end{equation}
where $\langle i,j \rangle$ indicates that the sum is over all the $j^{th}$ nearest neighbors of $S_{i}$, and where $H = [-\infty,\infty]$. The thermodynamic quantities can be obtained from the free energy $F$, which is defined as  $\beta F = -\ln(Z)$, where $\beta = (k_{B}T)^{-1}$, $k_{B}$ is the Boltzmann constant, $T$ is the temperature, and $Z$ is the canonical partition function. In this work, we set $k_{B} = 1$. The magnetization per spin $\mathcal{M}$ is defined as the average value of $S$ over all spins $\left\langle S \right\rangle $ and is found by taking the derivative of the free energy with respect to the external field at constant temperature, 
\begin{equation}
\mathcal{M} = -\left. \frac{\partial F}{\partial H} \right|_{T} = \left\langle S \right\rangle \;.
\label{eq:magnetization_theo}
\end{equation} 
Furthermore, the magnetic susceptibility per spin $\chi$ measures the change in the material's magnetization as the external field is varied and is obtained by taking the derivative of the magnetization with respect to the external field at constant temperature, 
\begin{equation}
\chi = \left. \frac{\partial \mathcal{M}}{\partial H} \right|_{T} = \beta \left\langle \Delta S^2 \right\rangle \;.
\label{eq:susceptibility_theo}
\end{equation} 
Moreover, the internal energy per spin $U$ measures the average energy of the system and is calculated by taking the derivative of $\beta F$ with respect to $\beta$ at constant external field, 
\begin{equation}
U = \left. \frac{\partial (\beta F)}{\partial \beta} \right|_{H} = \left\langle E \right\rangle\;.
\label{eq:meanEnergy_theo}
\end{equation}
Finally, the specific heat $C$ measures the change in the average energy (absorbed or released) as the temperature is varied and is obtained by the derivative of the average energy with respect to the temperature at constant external field, 
\begin{equation}
C = \left. \frac{\partial \langle E \rangle}{\partial T} \right|_{H} = - \beta^{2} \left.  \frac{\partial U}{\partial \beta} \right|_{H}  = \beta^{2}  \left\langle \Delta E^2 \right\rangle \;. 
\label{eq:specHeat_theo}
\end{equation}
The thermodynamic quantities may be derived analytically or computed from numerical simulations. 

\section{Derivation of the approximate thermodynamics of the 2D Ising model}

We derive in this section approximate expressions for the thermodynamic quantities associated with the two-dimensional Ising model in an external magnetic field by using the small angle approximation. We follow the Huang's interpretation \cite{re:huang87} of the Kaufman derivation \cite{re:kaufman49a}. 

\subsection{The quaternion representation of the 2D Ising model}

We first write the two-dimensional model in its quaternion representation. To do so, we consider a square lattice of spins of size $N = n \times n$. Assume that each spin can have states $S = \pm 1$. Let $\sigma^{k}$ be the set of all spins of the $k^{th}$ row, $\sigma^{k}= \{S_{1}, ... ,S_{n} \}^{k^{th}\,row}$. Assume isotropic nearest-neighbour interactions and periodic boundary conditions. Let $E(\sigma^{k},\sigma^{k+1})$ be the interaction energy between $\sigma^{k}$ and $\sigma^{k+1}$ and $E(\sigma^{k})$ be the interaction energy within the $k^{th}$ row. Assume further that an external magnetic field with strength $H$ is applied to the square lattice of spins. In these terms, the interaction energies are
\begin{equation}
E(\sigma^{k},\sigma^{k+1}) = -J\sum_{i=1}^{n} S_{i}^{k}S_{i}^{k+1} - \frac{H}{4}\sum_{i=1}^{n} \left(S_{i}^{k}+S_{i}^{k+1}\right)  \;,
\label{eq:energy_row}
\end{equation}
and 
\begin{equation}
E(\sigma^{k}) = -J\sum_{i=1}^{n} S_{i}^{k}S_{i+1}^{k} - \frac{H}{4}\sum_{i=1}^{n} \left(S_{i}^{k}+S_{i+1}^{k}\right) \;. 
\label{eq:energy_col}
\end{equation}
The partition function $Z$ is therefore given by
\begin{equation}
Z =  \sum_{\sigma^{1}}\cdots\sum_{\sigma^{n}} \prod_{k=1}^{n} \exp\left\{ -\beta \left[E(\sigma^{k},\sigma^{k+1})+ E(\sigma^{k}) \right]\right\} \;.
\end{equation}
Let $\mathcal{T}$ be the $2^{n} \times 2^{n}$ transfer matrix associated with $Z$ and defined by its matrix elements
\begin{equation}
\left\langle \sigma^{k} | \mathcal{T} | \sigma^{k+1} \right\rangle = \exp \left\{ -\beta \left[E(\sigma^{k},\sigma^{k+1})+ E(\sigma^{k})\right] \right\}\;.
\end{equation}
By using the identity $\sum_{\sigma^{k}} \left. | \sigma^{k} \right\rangle \left\langle \sigma^{k} | \right. = \mathds{1}$, the partition function reduces to 
\begin{equation}
Z = \sum_{\sigma^{1}}\left\langle \sigma^{1} | \mathcal{T}^{n} | \sigma^{1} \right\rangle  = \mathrm{Trace}[\mathcal{T}^{n}] \;.
\end{equation}
Let $\lambda_{1}$, ..., $\lambda_{2^{n}}$ be the $2^{n}$ eigenvalues of $\mathcal{T}$. Since the trace of a matrix is independent of its representation, the partition function is
\begin{equation}
Z = \sum_{k=1}^{2^{n}} (\lambda_{k})^{n} \;.
\end{equation}
If the eigenvalues of $\mathcal{T}$ are of order $e^{n}$, it is expected that
\begin{equation}
\lim_{N\rightarrow\infty} \frac{1}{N} \ln\left[ Z \right] = \lim_{n\rightarrow\infty} \frac{1}{n} \ln\left[ \lambda_{max} \right] \;,
\end{equation}
where $\lambda_{max}$ is the largest eigenvalue of $\mathcal{T}$. The problem thus reduces to solving for the largest eigenvalue for the transfer matrix $\mathcal{T}$. The matrix elements of $\mathcal{T}$ are defined as 
\begin{equation}
\langle S_{1}^{k},..., S_{n}^{k} | \mathcal{T} | S_{1}^{k+1}, ..., S_{n}^{k+1} \rangle = \prod_{i=1}^{n}  e^{\phi S_{i}^{k}S_{i+1}^{k}} e^{ \frac{\psi}{4} \left(S_{i}^{k} +S_{i+1}^{k} \right)} e^{\phi S_{i}^{k}S_{i}^{k+1}}e^{\frac{\psi}{4} \left(S_{i}^{k}+S_{i}^{k+1}\right) }\;,
\end{equation}
where $\phi = \beta J$ and $\psi = \beta H$. Define two $2^{n} \times 2^{n}$ matrices $V_{1}$ and $V_{2}$ with matrix elements given by
\begin{equation}
\langle S_{1}^{k}, ..., S_{n}^{k} | V_{1} | S_{1}^{k+1}, ..., S_{n}^{k+1} \rangle = \prod_{i=1}^{n} e^{\phi S_{i}^{k}S_{i}^{k+1}}e^{\frac{\psi}{4} \left(S_{i}^{k}+S_{i}^{k+1}\right)} \;,
\end{equation}
and
\begin{equation}
\langle S_{1}^{k}, ..., S_{n}^{k} | V_{2} | S_{1}^{k+1}, ..., S_{n}^{k+1} \rangle = \delta_{S_{1}^{k},S_{1}^{k+1}} ... \delta_{S_{n}^{k},S_{n}^{k+1}} \prod_{i=1}^{n} e^{\phi S_{i}^{k}S_{i+1}^{k}} e^{\frac{\psi}{4} \left(S_{i}^{k}+S_{i+1}^{k}\right)}  \;,
\end{equation}
where $\delta$ is the Kronecker delta symbol. The matrix $V_{2}$ is diagonal  in this representation. In these terms, the transfer matrix is given by
\begin{equation}
\mathcal{T} = V_{2}V_{1}\;.
\end{equation}
Our goal is to write the matrices $V_{1}$ and  $V_{2}$ in their quaternion representation. To do so, introduce three sets of $2^{n} \times 2^{n}$ matrices $X_{k}$, $Y_{k}$, and $Z_{k}$ ($k = 1, ..., n$) defined as
\begin{equation}
\begin{split}
X_{k} &= \mathds{1} \times \mathds{1} \times ... \times (X)_{k^{th}} \times ... \times \mathds{1} \;, \\
Y_{k} &= \mathds{1} \times \mathds{1} \times ... \times (Y)_{k^{th}} \times ... \times \mathds{1} \;, \\
Z_{k} &= \mathds{1} \times \mathds{1} \times ... \times (Z)_{k^{th}} \times ... \times \mathds{1} \;, 
\end{split}
\end{equation}
where 
\begin{equation} 
\mathds{1} = \begin{bmatrix} 1 & 0 \\ 0 & 1 \end{bmatrix}  ,  \quad X = \begin{bmatrix} 0 & 1 \\ 1 & 0 \end{bmatrix} \;, Y = \begin{bmatrix} 0 & -i \\ i & 0 \end{bmatrix}  ,  \quad Z = \begin{bmatrix} 1 & 0 \\ 0 & -1 \end{bmatrix} \;,
\label{eq:pauli}
\end{equation} 
are the Pauli matrices and are used as a basis in the quaternion representation. Consider first the matrix $V_{1}$, which is a direct product of $n$ $2 \times 2$ identical matrices, 
\begin{equation}
V_{1}= M \times M \times ... \times M \;,
\label{eq:V1_product}
\end{equation}
where $M$ is defined by 
\begin{equation}
M = \begin{pmatrix} e^{\phi+\psi/2} & e^{-\phi} \\ e^{-\phi} & e^{\phi-\psi/2} \end{pmatrix} = \cosh(\psi/2)e^{\phi}\mathds{1} +  \sinh(\psi/2)e^{\phi} Z + e^{-\phi} X  \;.
\label{eq:2d_transfer}
\end{equation}
Factorize the term proportional to $\mathds{1}$ and write Eq. (\ref{eq:2d_transfer}) as
\begin{equation}
M_{\pm} = \cosh(\psi/2)e^{\phi} \left[ \mathds{1} \pm \frac{\text{sec}(\varphi)}{\cosh(\psi/2)e^{2\phi}} X^{\prime}  \right]\;,
\label{eq:1d_transfer3}
\end{equation}
where we have defined 
\begin{equation}
\begin{bmatrix}  X^{\prime} \\ Z^{\prime} \end{bmatrix} = \begin{bmatrix} \cos(\varphi) & \sin(\varphi)  \\ -\sin(\varphi) & \cos(\varphi)  \end{bmatrix} \begin{bmatrix}  X \\ Z \end{bmatrix} \;,
\label{eq:rotFrame}
\end{equation} 
and
\begin{align}
\cos(\varphi) &= \frac{1}{\sqrt{1+\sinh^{2}(\psi/2)e^{4\phi}}} \;, \label{eq:cos_varphi} \\
\sin(\varphi) &= \frac{\sinh(\psi/2)e^{2\phi}}{\sqrt{1+\sinh^{2}(\psi/2)e^{4\phi}}} \label{eq:sin_varphi}\;,
\end{align} 
and $\tan(\varphi)$ = $\sinh(\psi/2)e^{2\phi}$. Here, $M_{\pm}$ represents the two equivalent solutions corresponding to the two choices of sign in front of $X^{\prime}$; because $(X^{\prime})^2 = \mathds{1}$, both choices lead to the same eigenvalues. The angle $\varphi$ represents a rotation between the reference frame $(X,Z)$ and a rotated frame $(X^{\prime},Z^{\prime})$. In these terms, the matrix $M_{\pm}$ reduces to 
\begin{equation}
M_{\pm} = \cosh(\psi/2)e^{\phi} \left[\mathds{1} \pm \tanh(\omega) X^{\prime}\right]\;,
\label{eq:1d_transfer5}
\end{equation}
where we have defined
\begin{equation}
\tanh(\omega) = \frac{\text{sec}(\varphi)}{\cosh(\psi/2)e^{2\phi}} \;.
\end{equation}
By using the exponential matrix definition $e^{b A} = \cosh(b)+\sinh(b)A$, where $A$ is a matrix satisfying $A^2 = 1$, and where $b$ is a constant, Eq. (\ref{eq:1d_transfer5}) can be written as 
\begin{equation}
M_{\pm} = \frac{\cosh(\psi/2)e^{\phi}}{\cosh(\omega)} \exp \left( \pm \omega X^{\prime}   \right) \;.
\label{eq:Mpm_rotation}
\end{equation}
By performing the product defined in Eq. (\ref{eq:V1_product}), we obtain
\begin{equation}
V_{1} = \left[\frac{\cosh(\psi/2) e^{\phi}}{\cosh{\omega}} \right]^n \prod_{k=1}^{n}  e ^{\pm \omega X_{k}^{\prime} } \;.
\end{equation}
Let's now consider $V_{2}$. By using the quaternion basis, the matrix is 
\begin{equation}
V_{2} =  \prod_{k=1}^{n} e^{\phi Z_{k}Z_{k+1}} e^{\frac{\psi}{4} \left( Z_{k}+Z_{k+1} \right)}  \;.
\end{equation}
Use the following identities
\begin{equation}
e^{\phi Z_{k}Z_{k+1}} =  \cosh(\phi)\mathds{1} + \sinh(\phi)Z_{k}Z_{k+1}  \;,
\end{equation}
and
\begin{equation}
e^{\frac{\psi}{4} \left( Z_{k}+Z_{k+1} \right)} = \cosh^{2}(\psi/4)\mathds{1}+ \sinh^{2}(\psi/4)Z_{k}Z_{k+1}   +\frac{1}{2}\sinh(\psi/2)\left( Z_{k}+Z_{k+1} \right)  \;,
\end{equation}
and perform the multiplication to obtain 
\begin{equation}
V_{2} = \left[\frac{e^{-\phi}}{2}\right]^n \prod_{k=1}^{n} \left[\mu\mathds{1}+\nu Z_{k}Z_{k+1}  +\tan(\varphi)\left(Z_{k}+Z_{k+1} \right) \right] \;,
\end{equation}
where
\begin{align}
\mu &= \cosh(\psi/2)e^{2\phi}+1  \;,\\
\nu &= \cosh(\psi/2)e^{2\phi}-1  \;.
\end{align}
By combining results from $V_{1}$ and $V_{2}$, the transfer matrix is  
\begin{equation}
\mathcal{T} = \left[\frac{\cosh(\psi/2) }{2\cosh(\omega)} \right]^n \prod_{k=1}^{n} e ^{ \pm\omega X_{k}^{\prime} } \prod_{k=1}^{n} \left[\mu\mathds{1}+\nu Z_{k}Z_{k+1} +\tan(\varphi)\left(Z_{k}+Z_{k+1} \right) \right] \;.
\label{eq:transfer_full}
\end{equation}

\subsection{Small-angle approximation applied to the 2D Ising model}

We now apply the small-angle approximation to Eq. (\ref{eq:transfer_full}). Let $\varphi \to 0$, so that $X_{k}^{\prime} = X_{k}$ and $\tan(\varphi) = 0$. In the exact lattice Hamiltonian, the field term $-H\sum_i S_i$ is distributed as four contributions of strength $H/4$ around each spin. After the quaternion rotation and the small-angle approximation, the effective transfer matrix still couples to this per-bond field. To ensure that the system responds to the full physical field $H$ in the limit $H \to 0$, we rescale $\psi \to 4\psi$ in Eq. (\ref{eq:transfer_full}). In these terms, the small-angle approximation is understood as
\begin{equation}
\varphi \to 0 \Rightarrow \varphi_{a} = \lim_{\begin{subarray}{l}
\psi \to 0 \\
\phi \to 0
\end{subarray}} \text{arctan}\left[ \sinh(2\psi)e^{2\phi} \right] \;.
\end{equation}
where $a$ stands for approximate. The approximation is valid as $\varphi_{a}$ is small which implies that $2\phi = 2\beta J \ll 1$ and $2\psi = 2\beta H \ll 1$. These large temperatures satisfy $T \gg 2(H+J)$ which is obtained by expanding the condition $\varphi_{a} \ll 1$. Under this approximation, the transfer matrix $\mathcal{T}_{a}$ reduces to
\begin{equation}
\mathcal{T}_{a} = \left[\frac{\cosh(2\psi) }{2\cosh(\omega_{a})} \right]^n \prod_{k=1}^{n} e ^{ \omega_{a} X_{k} } \prod_{k=1}^{n}  \bigg\{ [\cosh(2\psi)e^{2\phi}+1] \mathds{1} + \left[\cosh(2\psi)e^{2\phi}-1\right] Z_{k}Z_{k+1}  \bigg\} \;,
\end{equation}
where 
\begin{equation}
\coth(\omega_{a}) = \cosh(2\psi)e^{2\phi} \;.
\end{equation}
By using the definition of the matrix exponential, the transfer matrix simplifies to
\begin{equation}
\mathcal{T}_{a} = \left[\frac{\cosh(2\psi)e^{\phi} }{\cosh(\omega_{a})} \right]^n \left[\frac{\cosh(2\psi)e^{\phi}+e^{-\phi} }{2\cosh(\eta)} \right]^n \prod_{k=1}^{n} e^{ \omega_{a}  X_{k}} \prod_{k=1}^{n} e^{ \eta Z_{k}Z_{k+1} } \;,
\label{eq:tranfer_2d}
\end{equation}
where we have defined
\begin{equation}
\tanh(\eta) = \frac{\cosh(2\psi)e^{2\phi}-1}{\cosh(2\psi)e^{2\phi}+1} \;.
\end{equation}
The factors of Eq. (\ref{eq:tranfer_2d}) can be further simplified by using relations introduced in Appendix \ref{app:omega_eta}, namely
\begin{equation}
\frac{\cosh(2\psi)e^{\phi} }{\cosh(\omega_{a})} = e^{-\phi} \sqrt{\cosh^2(2\psi)e^{4\phi}-1} = \sqrt{2\sinh(2\eta)\cosh(2\psi)} \;,
 \label{eq:f1}
\end{equation}
and 
\begin{equation}
\frac{\cosh(2\psi)e^{\phi}+e^{-\phi} }{2\cosh(\eta)} = \sqrt{\cosh(2\psi)} \;. 
\label{eq:f2}
\end{equation}
By substituting Eqs. (\ref{eq:f1}) and (\ref{eq:f2}) into Eq. (\ref{eq:tranfer_2d}) , the transfer matrix is
\begin{equation}
\mathcal{T}_{a} = \left[2\sinh(2\eta) \cosh^2(2\psi) \right]^\frac{n}{2} \prod_{k=1}^{n} e^{ \omega_{a}  X_{k}} \prod_{k=1}^{n} e^{ \eta Z_{k}Z_{k+1} } \;.
\label{eq:tranfer_2d_10}
\end{equation}
It is worth noting the progress made so far. The small-angle approximation allowed the reduction of the transfer matrix in the same form as the transfer matrix of the two-dimensional model with $H = 0$. In this form, the dependence on the external field strength is included in the parameters $\omega_{a}$ and $\eta$, and in the first term factoring the product. The approximation therefore allows using earlier tools developed by Kaufman \cite{re:kaufman49a} to find the largest eigenvalue. In particular, we further note the following identities
\begin{equation}
\tanh(\eta) = e^{-2\omega_{a}}  \;,
\end{equation}
and
\begin{equation}
\sinh(2\omega_{a})\sinh(2\eta) = 1 \;.
\label{eq:sinesine}
\end{equation}
The largest eigenvalue is found by writing $\mathcal{T}_{a}$ in terms of spins representative of plane rotation. To do so, introduce $2n$ matrices of size $2^n \times 2^n$ satisfying the anticommutation rule $\Gamma_{p}\Gamma_{q}+\Gamma_{q}\Gamma_{p} = 2\delta_{pq}$, where $\delta_{pq} = 1$  if $p=q$ and $\delta_{pq} = 0$ otherwise. Let $\Gamma_{2t-1} = \prod_{s=1}^{t-1}X_{s}Z_{t}$ and $\Gamma_{2t} = \prod_{s=1}^{t-1}X_{s}Y_{t}$ with $t = \{1,...,n\}$. Under this representation, the matrices $V_{1}$ and $V_{2}$ are expressed as spin representative of a product of commuting plane rotations for which the eigenvalues are known. By applying this machinery, the approximate free energy is 
\begin{equation}
\beta F_{a} = -\frac{1}{2}\log\left[ 2\sinh(2\eta)\cosh^2(2\psi) \right] - I \;,
\end{equation}
where $I$ is an integral defined as 
\begin{equation}
I = \frac{1}{2\pi^{2}} \int_{0}^{\pi}\int_{0}^{\pi} d\upsilon d\upsilon^{\prime} \log\left[ 2\cosh(2\omega_{a})\cosh(2\eta)  - 2\cos(\upsilon)-2\cos(\upsilon^{\prime}) \right]  \;,
\label{eq:integral}
\end{equation}
where we have used Eq. (\ref{eq:sinesine}). Equation (\ref{eq:integral}) can be simplified in terms of an elliptic integral. The steps involved in the reduction of the integral are shown in Appendix \ref{app:reduction}. By using the identity \cite{re:onsager44}
\begin{equation}
|z| = \frac{1}{\pi} \int_{0}^{\pi} d\tau \log\left[ 2\cosh(z)-2\cos(\tau) \right] \;,
\end{equation}
the integral is reduced to 
\begin{equation}
I = \frac{1}{2}\log\left[\frac{2}{\kappa}\right] +\frac{1}{\pi} \int_{0}^{\pi/2} d\theta \log \left[1+\sqrt{1-\kappa^{2}\sin^{2}(\theta)}\right] \;,
\end{equation}
where $\kappa$ is named the elliptic modulus and is defined by
\begin{equation}
\kappa = \frac{2\sinh(2\eta)}{\cosh^{2}(2\eta)} \;. 
\label{eq:kappa}
\end{equation}
Therefore, the free energy of the two-dimensional Ising model in an external magnetic field in the small-angle approximation is 
\begin{equation}
\beta F_{a} = -\log\left[ 2\cosh(2\eta)\cosh(2\psi) \right] -\frac{1}{\pi} \int_{0}^{\pi/2} d\theta \log\left\{\frac{1}{2}  \left[1+\sqrt{1-\kappa^{2}\sin^{2}(\theta)}\right]\right\}\;.
\label{eq:free_2D}
\end{equation}
Note that as $H = 0$, then  $\eta = \beta J$, and Eq. (\ref{eq:free_2D}) is identical to the free energy of the two-dimensional Ising model with $H = 0$. 

\subsection{Approximate thermodynamics of the 2D Ising model}

The approximate thermodynamic quantities can be derived from Eq. (\ref{eq:free_2D}). Consider the mathematical identities introduced in Appendix \ref{app:identities}. The magnetization is obtained by taking the derivative of the free energy with respect to $H$. However, as a consequence of the small-angle approximation where we let $\psi \to 4\psi$, the magnetization must be corrected by a factor of 1/4, which leads to
\begin{equation}
\mathcal{M}_{a} = \frac{1}{4} \mathcal{M}^{\prime} = \frac{1}{2}\tanh(2\psi) \left\{ 1+\frac{1}{2}\text{coth}(2\eta) \left[1+\frac{2}{\pi}\kappa^{\prime}K(\kappa)\right]  \right\}  \;,
\label{eq:ma_2d}
\end{equation}
where $K$ is the complete elliptic integral of the first kind
\begin{equation}
K(\kappa) = \displaystyle{\int_{0}^{\pi/2}\frac{d\theta}{\sqrt{1-\kappa^{2}\sin^{2}(\theta)}}} \;,
\end{equation}
and where 
\begin{equation}
\kappa^{\prime} = 2\tanh^{2}(2\eta)-1 \;.
\end{equation}	
Furthermore, the susceptibility $\chi_{a} $ is obtained by taking the derivative of $\mathcal{M}_{a}$ with respect to $H$, which leads to 
\begin{equation}
\begin{split}
\chi_{a} &= \beta \text{sech}^2(2\psi) \left\{ 1+\frac{1}{2}\text{coth}(2\eta)\Big[1+\frac{2}{\pi}\kappa^{\prime}K(\kappa)\Big] \right\} \\
&+\frac{\beta}{2\pi}\tanh^{2}(2\psi)\coth^2(2\eta) \left\{2\Big[K(\kappa)-E(\kappa)\Big]-(1-\kappa^{\prime}) \Big[\frac{\pi}{2}+\kappa^{\prime}K(\kappa)\Big]   \right\} \;, \\
\end{split}
\label{eq:chia_2d}
\end{equation}
where $E(\kappa)$ is the complete elliptic integral of the second kind
\begin{equation}
E(\kappa) = \int_{0}^{\pi/2} d\theta \sqrt{ 1 - \kappa^{2}\sin^{2}(\theta)} \;.
\end{equation}
Note that Eq. (\ref{eq:chia_2d}) could also have been obtained by taking the second derivative of the free energy with respect to $H$, and correcting the result by a factor of 1/4. Moreover, the average energy $U_{a}$ is obtained by taking the derivative of the free energy with respect to $\beta$, and correcting the magnetization by $1/4$ so that 
\begin{equation}
U_{a} = -J \text{coth}(2\eta)\left[1+\frac{2}{\pi}\kappa^{\prime}K(\kappa)\right]-H\mathcal{M}_{a} \;.
\label{eq:ua_2d}
\end{equation}
Finally, the specific heat is obtained by taking the derivative of $U_{a}$ with respect to $\beta$ and multiplying by $-\beta^2$, which leads to
\begin{equation}
\begin{split}
C_{a} &= \frac{2}{\pi} \beta^{2}\text{coth}^{2}(2\eta) \left(\frac{\partial \eta}{\partial \beta}\right)^2 \bigg\{ 2\Big[K(\kappa)-E(\kappa)\Big]-(1-\kappa^{\prime})\left[\frac{\pi}{2}+\kappa^{\prime}K(\kappa) \right] \bigg\} \\
&- \frac{3}{2\pi} \beta\psi\tanh(2\psi)\text{coth}^{2}(2\eta) \frac{\partial \eta}{\partial \beta}  \bigg\{ 2\Big[K(\kappa)-E(\kappa)\Big]-(1-\kappa^{\prime})\left[\frac{\pi}{2}+\kappa^{\prime}K(\kappa) \right] \bigg\}\\
&+\psi^2 \text{sech}^{2}(2\psi) \left\{ 1+\frac{1}{2}\text{coth}(2\eta)\Big[  1+\frac{2}{\pi}\kappa^{\prime}K(\kappa) \Big] \right\} \;,
\end{split}
\label{eq:ca_2d}
\end{equation}
where
\begin{equation}
\frac{\partial \eta}{\partial \beta} = J+H \tanh(2\beta H) \;.
\end{equation}

By setting $H = 0$, Eqs. (\ref{eq:ua_2d}) and (\ref{eq:ca_2d}) reduce to the exact internal energy per spin and the specific heat of the two-dimensional Ising model with $H = 0$. In the same settings, the magnetization vanishes $\mathcal{M}_{a} = 0$, but the susceptibility is 
\begin{equation}
\chi_{a}^{H=0} = \beta\left\{ 1+\frac{1}{2}\text{coth}(2\beta J)\left[1+\frac{2}{\pi}\kappa^{\prime}K(\kappa)\right] \right\}  \;, \\
\label{eq:chia_2d_H0}
\end{equation}
where $\kappa^{\prime} = 2\tanh^{2}(2\beta J)-1$. Equation (\ref{eq:chia_2d_H0}) does not capture the dynamics of the model as $H = 0$ near the critical temperature, namely a divergence at $T_{c}$.

\section{Numerical simulations with the BKL algorithm} \label{sec:simulation}

The two-dimensional Ising model was simulated with the BKL algorithm \cite{re:bkl75}. When the system is close to equilibrium or in a metastable state, the rate of generating new configurations is slow because transition probabilities are small. The BKL algorithm speeds up the calculation by performing a transition at each iteration proportionally to the probability of occurrence of this transition. A quantity representing the simulation time is updated accordingly. The steps involved in the update of a configuration and the transition probabilities used are shown in Table \ref{tab:transition}. All spins were randomly initialized in $S = -1$ or $S = 1$. Their classes were determined and the number of spins in each class was updated accordingly. A series of iterations over time was first executed to thermalize the system up to a predefined time ($\Delta t_{therma}$). After thermalization, iterations were conducted to record independent configurations of the system's state. In each iteration, BKL sampling was performed up to a predefined time ($\Delta t_{compute}$). After these lattice updates, the number of spins in each class was recorded. The thermodynamic quantities were computed after the simulation from the recorded number of spins in each class. Let  $\mathcal{S}$ be the sum of all spin states of a configuration normalized by the total number of spins computed as 
\begin{equation}
\mathcal{S} = \frac{1}{n^2}\sum_{i,j=1}^{n}S_{i,j} \;.
\end{equation}		
The magnetization $\mathcal{M}$ and susceptibility $\chi$ were computed by calculating the average and variance of $\mathcal{S}$ over all configurations such that
$\mathcal{M} = \langle \mathcal{S} \rangle$ and $\chi = \beta\langle \Delta \mathcal{S}^2 \rangle$. Furthermore, let $\mathcal{E}$ be the energy per spin of a configuration computed as 
\begin{equation}
\begin{split}
\mathcal{E} =& -\frac{J}{2n^2}\sum_{i,j=1}^{n} \left(S_{i,j}S_{i-1,j}+S_{i,j}S_{i+1,j} + S_{i,j}S_{i,j-1}+S_{i,j}S_{i,j+1}\right) -\frac{H}{n^2}\sum_{i,j=1}^{n}S_{i,j}\;,
\end{split}
\end{equation}		
where a factor of 1/2 was applied to account for double counting. The internal energy $U$ and the specific heat $C$ were computed with $U = \langle \mathcal{E} \rangle$ and $C = \beta^{2}\langle \Delta \mathcal{E}^2 \rangle$. Square lattices of size $N = n \times n$, with $n=\{20, 50\}$ were used with parameters $\Delta t_{therma}$ = 5 $\times$ 10$^{4}$ and $\Delta t_{compute}$ = 250. A total of 100,800 configurations were simulated. We validated our algorithm by considering the independent cases where $H = 0$ and $J = 0$. These results are shown in the supplementary material.

\section{Comparison between analytical and numerical results} \label{sec:comparison}

The approximate thermodynamic expressions [Eqs. (\ref{eq:ma_2d}), (\ref{eq:chia_2d}), (\ref{eq:ua_2d}), and (\ref{eq:ca_2d})] are compared with numerical simulations as the angle $\varphi_{a}$ is varied in Fig. \ref{fig:varphi_2D}. The cases where $J > H$, $J = H$, and $J < H$ were considered. For an $N = 50 \times 50$ lattice size, the absolute difference between the approximate and numerical values was lower than 1 in all approximate thermodynamic quantities as $\varphi_{a} < 0.01$ at $J = 0.01$ and $H = 0.001$, $\varphi_{a} < 0.1$ at $J = 0.01$ and $H = 0.01$, and $\varphi_{a} < 0.03$ at $J = 0.01$ and $H = 0.0001$. These small differences were obtained in regions where temperatures were larger than the critical temperature ($\beta_{c}J = \ln[\sqrt{2}+1]/2$).

The approximate thermodynamic expressions are further compared with numerical simulations as the external field strength is varied in Fig. \ref{fig:thermo2D_J001} at $J$ = 0.01 and $T = [0.4, 1, 2]  > T_{c}$. By considering numerical results obtained with $N = 50 \times 50$, the absolute difference between the approximate and numerical values was lower than 0.22 in $H = [-0.1,0.1]$ with $T = 0.4$, 0.03 in $H = [-0.25,0.25]$ with $T = 1$, and 0.01 in $H = [-0.5,0.5]$ with $T = 2$ in all thermodynamic quantities. As expected, the interval of magnetic field strength where the approximate expressions and the numerical values agree increased with increasing temperature. It is worth noting that $|H| > J$ in almost all results shown in this figure.

The approximate expression for the susceptibility at $H = 0$ [Eq. (\ref{eq:chia_2d_H0})] is shown in Fig. \ref{fig:susceptibility_H0} at $J = [0.01,1]$. The absolute difference between the approximate and numerical values decreased with increased temperature at both interaction energies. Equation (\ref{eq:chia_2d_H0}) is further compared with the susceptibility obtained with the mean-field theory and a perturbative approach at high temperature \cite{re:Orrick01b}. In the mean-field theory, the susceptibility $\chi^{MF}$ is 
\begin{equation}
\chi^{MF} = \frac{\beta}{\cosh^2\left(z\beta J \mathcal{M}^{MF} \right)-z\beta J} \;,
\end{equation}
where the magnetization $\mathcal{M}^{MF}$ is obtained from the self-consistent relation
\begin{equation}
\mathcal{M}^{MF} = \tanh\left(z\beta J \mathcal{M}^{MF} \right) \;,
\end{equation}
and where $z = 4$ is the coordination number. Furthermore, following a perturbative approach at high temperature, the susceptibility $\chi^{perturbative}$ can be approximated by
\begin{equation}
\chi^{perturbative} =  \beta \big(1-s^2\big)^{1/4} \big(1-s^{1/2}\big)^{-2} \;,
\end{equation}
where $s = \sinh^{2}(2\beta J)$. At the parameters considered, the perturbative approach provided greater accuracy than the other approximations. However, the small-angle approximation also provided accurate results. In both cases, the absolute difference between the approximate and numerical values was smaller than 0.2 for reduced temperatures larger than $T/J = 6$ at $J = 1$ and smaller than 1 for reduced temperatures larger than $T/J = 20$ at $J = 0.01$.

\section{Discussion}

By using the quaternion algebra, we introduced a small-angle approximation between a rotated and reference frame that allowed the derivation of approximate expressions for the thermodynamics of the two-dimensional Ising model in an external magnetic field. The analytical expressions were compared with numerical simulations of the exact model performed with the BKL algorithm. The agreement between the approximate expressions and numerical simulations was excellent in regions where the approximation is valid, namely at small $\phi$ and $\psi$. This work is the first comprehensive study providing expressions for the thermodynamics of the system that include the dependence on the external field and reduce to the known expressions as $H = 0$.

The approximate thermodynamic expressions have a singularity at a temperature defined by $\kappa^{\prime} = 0$. This singularity predicts a sharp transition between the ordered and disordered phase as shown in Fig. \ref{fig:thermo2D_J001}, which can be observed as a divergence in $\chi$ and $C$. However, the transition is not sharp as $H \neq 0$, as shown with the numerical simulations. This disagreement was expected as the approximation is not valid in those regions. Furthermore, the approximate susceptibility as $H = 0$ is smooth around $\beta_{c} J$ but diverges in the numerical simulations. This was also expected as the small-angle approximation is not valid near $\beta_{c} J$, which further illustrates the limit of the approximation's applicability.

Only the two-dimensional isotropic case was considered in this work, namely the interaction energy $J$ was the same for all nearest neighbors. Due to the form of the transfer matrix after the small-angle approximation, the anisotropic case could be derived with methods introduced previously \cite{re:mccoy73,re:baxter82}. Furthermore, approximate expressions for the spin-spin correlation function may be derived \cite{re:wu76,re:kaufman49b}. The approximate thermodynamics might also be derived with other lattice types in two dimensions, such as triangular or hexagonal lattices \cite{re:houtappel50}. Moreover, the approximation may be applied to models with finite size \cite{re:hucht17}. The rotated frame allowed the small-angle approximation to arise naturally in the quaternion representation of the model. Development of this framework in other representations, such as the free-fermion representation \cite{re:schultz64} or the dimer representation \cite{re:kasteleyn61}, may provide further insights into the derivation of the full model. However, attempts to describe the spontaneous magnetization of the model with $H = 0$ \cite{re:yang52,re:baxter11} may fail as the approximation introduced in this work is not valid around the critical temperature.

In this work, the external magnetic field strength was rescaled by a factor of 4 in two dimensions and by a factor of 2 in one dimension (see the derivation of the small-angle approximation applied to the one-dimensional Ising model in the supplementary material). These factors were necessary for the approximate thermodynamic expressions to agree with the numerical simulations. Interestingly, these factors led to the same definition of the angle between the rotated and reference frame in one and two dimensions, $\varphi_{a} = \text{arctan}[\sinh(2\beta H)\exp(2\beta J)]$. Future investigations to determine if this definition holds in lattices with different coordination numbers would be interesting.

In the small-angle approximation, the transfer matrix is centrosymmetric and methods developed in the solution of the two-dimensional Ising model with $H = 0$ can be applied. The exact solution of the model in an external magnetic field requires an approach able to deal with non-centrosymmetric matrix. The fundamental picture used in this work is geometrical; the Ising model is intrinsically related to rotations in a d-dimensional space. Considering not only rotations but also other geometrical transformations may help progress toward the exact solution. The three-dimensional Ising model is also unsolved. As such, it is unknown if the model can be written as a quaternion in three dimensions. However, if this is the case, we speculate that the machinery developed in this work would apply to the three-dimensional Ising model in an external field. 

\section{Conclusions}

Approximate expressions for the thermodynamics of the two-dimensional Ising model in an external magnetic field are provided. The analytical approximate expressions and numerical simulations agree in regions where the approximation is valid.


\section{Acknowledgments}

This work was produced with support from the University of Wollongong / National Computational Infrastructure Partner Share Scheme at the University of Wollongong. 


\newpage
\appendix

\section{Relations involving $\omega_a$ and $\eta$}{\label{app:omega_eta}}

In this appendix, we introduce useful relations involving $\omega_a$ and $\eta$. Consider first the definition of $\omega_{a}$,
\begin{equation} 
\omega_{a} = \text{arctanh}\left[\frac{1}{\cosh(2\psi)e^{2\phi}}\right] = \frac{1}{2} \ln\left[\frac{\cosh(2\psi)e^{2\phi}+1}{\cosh(2\psi)e^{2\phi}-1}\right] \;,
\end{equation}
and use $\cosh[\text{arctanh}(x)] = (1-x^{2})^{-1/2}$ and $\sinh[\text{arctanh}(x)] = x(1-x^{2})^{-1/2}$ to derive the following identities
\begin{align}
 \sinh(\omega_{a}) &= \frac{1}{\sqrt{\cosh^2(2\psi)e^{4\phi}-1}}  \;,\\
\cosh(\omega_{a}) &= \frac{\cosh(2\psi)e^{2\phi}}{\sqrt{\cosh^2(2\psi)e^{4\phi}-1}}   \;,\\
\sinh(2\omega_{a}) &= \frac{2\cosh(2\psi)e^{2\phi}}{\cosh^2(2\psi)e^{4\phi}-1}  \;,\\
\cosh(2\omega_{a}) &= \frac{\cosh^2(2\psi)e^{4\phi}+1}{\cosh^2(2\psi)e^{4\phi}-1}  \;.
\end{align}
Repeat the same exercise with the definition of $\eta$,
\begin{equation}
\eta =  \text{arctanh}\left[\frac{\cosh(2\psi)e^{2\phi}-1}{\cosh(2\psi)e^{2\phi}+1}\right] = \frac{1}{2} \ln\left[\cosh(2\psi)e^{2\phi}\right] \;,
\end{equation}
to derive
\begin{align}
\sinh(\eta) &= \frac{\cosh(2\psi)e^{\phi}-e^{-\phi}}{2 \sqrt{\cosh(2\psi)} }   \;,\\
\cosh(\eta) &= \frac{\cosh(2\psi)e^{\phi}+e^{-\phi}}{2 \sqrt{\cosh(2\psi)} }   \;,\\
\sinh(2\eta) &= \frac{ \cosh^2(2\psi)e^{4\phi}-1 }{ 2e^{2\phi}\cosh(2\psi) }  \;,\\
\cosh(2\eta) &= \frac{ \cosh^2(2\psi)e^{4\phi}+1 }{ 2e^{2\phi}\cosh(2\psi) }  \;.
\end{align}
Furthermore, we note the following relationships between $\omega_a$ and $\eta$,
\begin{align}
\tanh(\eta) &= e^{-2\omega_{a}}  \;,\\
\sinh(2\omega_{a})\sinh(2\eta) &= 1 \;,\\
\cosh(2\omega_{a})\tanh(2\eta) &= 1 \;.
\end{align}


\section{Reduction of the integral}{\label{app:reduction}}

In this appendix, we reduce the integral
\begin{equation}
I = \frac{1}{2\pi^{2}} \int_{0}^{\pi}\int_{0}^{\pi} d\upsilon d\upsilon^{\prime} \log\left\{ 2\cosh(2\omega)\cosh(2\eta)  - 2 [\cos(\upsilon)+\cos(\upsilon^{\prime})] \right\} \;.
\end{equation}
Introduce the change of variables $\upsilon = \theta_{1}+\theta_{2}/2$ and $\upsilon^{\prime} = \theta_{1}-\theta_{2}/2$, where $0 \leq \upsilon \leq \pi$ and $0 \leq 	\upsilon^{\prime} \leq \pi$, so that $0 \leq \theta_{1} \leq \pi$ and $0 \leq \theta_{2} \leq \pi$ and its associated Jacobian
\begin{equation}
\begin{vmatrix} \frac{\partial \upsilon^{\prime}}{\partial \theta_{1}} & \frac{\partial \upsilon^{\prime}}{\partial \theta_{2}} \\ \frac{\partial \upsilon}{\partial \theta_{1}} & \frac{\partial \upsilon}{\partial \theta_{2}} \end{vmatrix} = \begin{vmatrix} 1 & -1/2 \\ 1 & 1/2 \end{vmatrix} = 1 \;.
\end{equation}
Then, upon the further substitution $\theta_{2} \rightarrow 2\theta_{2}$ (so that $0 \leq \theta_{2} \leq \pi/2$), the integral becomes
\begin{equation}
I = \frac{1}{\pi^{2}} \int_{0}^{\pi}\int_{0}^{\pi/2}  d\theta_{1} d\theta_{2} \log\left[ 2\cosh(2\omega)\cosh(2\eta) - 4\cos(\theta_{1})\cos(\theta_{2}) \right] \;,
\end{equation}
which can be written as
\begin{equation}
\begin{split}
I = \frac{1}{\pi^{2}} \int_{0}^{\pi}\int_{0}^{\pi/2}  d\theta_{1} d\theta_{2} \bigg\{  & \log\left[2\cos(\theta_{2})\right] \bigg. \\
&+ \bigg. \log\left[ \frac{2\cosh(2\omega)\cosh(2\eta)}{2\cos(\theta_{2})}- 2\cos(\theta_{1}) \right]   \bigg\} \;.
\end{split}
\end{equation}
Now use the identity
\begin{equation}
|z| = \frac{1}{\pi} \int_{0}^{\pi} d\tau \log\left[ 2\cosh(z)-2\cos(\tau) \right] \;,
\end{equation}
and set 
\begin{equation}
\cosh(z) = \frac{\cosh(2\omega)\cosh(2\eta)}{2\cos(\theta_{2})} \;,
\end{equation}
to reduce the double integral to a single integral, 
\begin{equation}
I = \frac{1}{\pi} \int_{0}^{\pi/2}  d\theta_{2}   \log\left[2\cos(\theta_{2})\right] + \frac{1}{\pi} \int_{0}^{\pi/2}   d\theta_{2}  \text{acosh} \left[\frac{\cosh(2\omega)\cosh(2\eta)}{2\cos(\theta_{2})}\right]   \;.
\end{equation}
Now define the elliptic modulus
\begin{equation}
\kappa = \frac{2}{\cosh(2\omega)\cosh(2\eta)} = \frac{2\sinh(2\eta)}{\cosh^2(2\eta)} \;.
\end{equation}
In these terms, the integral is
\begin{equation}
I = \frac{1}{\pi} \int_{0}^{\pi/2}   d\theta_{2} \log\left[ 2\cos(\theta_{2})\right]+\frac{1}{\pi} \int_{0}^{\pi/2}   d\theta_{2} \text{acosh} \left[ \frac{1}{\kappa \cos(\theta_{2})} \right] \;.
\end{equation}
Use $ \text{acosh}(x) = \log(x+\sqrt{x^{2}-1})$ to write
\begin{equation}
\begin{split}
I &= \frac{1}{\pi} \int_{0}^{\pi/2}   d\theta_{2} \log\left[ 2\cos(\theta_{2})\right] \\
&+\frac{1}{\pi} \int_{0}^{\pi/2}   d\theta_{2}\log\left\{ \left[\frac{1}{2\cos(\theta_{2})}\right]\left[\frac{2}{\kappa} \right] \bigg[1+\sqrt{1-\kappa^{2} \cos^2(\theta_{2})}\right] \bigg\} \;.
\end{split}
\label{eq:I_temp}
\end{equation}
Expand the logarithmic term of the second term of Eq. (\ref{eq:I_temp}) to cancel the first term of the same equation. Now let $\theta_{2} = \pi/2-\theta$ so that $\cos(\theta_{2}) = \sin(\theta)$ to write the integral as
\begin{equation}
I = \frac{1}{2}\log\left[\frac{2}{\kappa}\right] +\frac{1}{\pi} \int_{0}^{\pi/2} d\theta \log \left[1+\sqrt{1-\kappa^{2}\sin^{2}(\theta)}\right] \;.
\label{eq:I_complete}
\end{equation}
Equation (\ref{eq:I_complete}) completes the reduction of the integral $I$. Now consider taking the derivative of $I$ with respect to an arbitrary variable $\xi$
\begin{equation}
\frac{\partial I}{\partial \xi} = -\frac{1}{2\kappa} \frac{\partial \kappa}{\partial \xi} - \frac{1}{\pi}\int_{0}^{\pi/2} d\theta  \frac{\kappa\sin^{2}(\theta)}{\Delta(1+\Delta)}\frac{\partial \kappa}{\partial \xi} \;,
\end{equation}
where $\Delta = \sqrt{1-\kappa^{2}\sin^{2}(\theta)}$. Using $(1+\Delta)(1-\Delta)=\kappa^{2}\sin^{2}(\theta)$, the derivative is
\begin{equation}
\frac{\partial I}{\partial \xi} = -\frac{1}{2\kappa} \frac{\partial \kappa}{\partial \xi} - \frac{1}{\pi} \frac{1}{\kappa} \frac{\partial \kappa}{\partial \xi} \left\{ -\frac{\pi}{2} + K(\kappa) \right\} =  -\frac{1}{\pi} \frac{1}{\kappa} \frac{\partial \kappa}{\partial \xi} K(\kappa) \;,
\end{equation}
where $K(\kappa)$ is the complete elliptic integral of the first kind
\begin{equation}
K(\kappa) = \displaystyle{\int_{0}^{\pi/2}\frac{d\theta}{\sqrt{1-\kappa^{2}\sin^{2}(\theta)}}} \;.
\end{equation}


\section{Identities to calculate the thermodynamics} {\label{app:identities}}

In this appendix, we introduce identities that are useful in the derivation of the approximate expressions of the system's thermodynamics. Consider the elliptic integral of the first kind \cite{re:abramowitz65}, 
\begin{equation}
K(\kappa) = \int_{0}^{\pi/2} \frac{d\theta}{\sqrt{1-\kappa^{2}\sin^{2}(\theta)}} \;,
\end{equation}
where $\kappa$ is the elliptic modulus. The derivative of $K(\kappa)$ with respect to $\kappa$ is 
\begin{equation}
\frac{\partial K}{\partial \kappa} = \frac{1}{\kappa} \left[\frac{E(\kappa)}{1-\kappa^2} - K(\kappa) \right] \;,
\end{equation}
where $E(\kappa)$ is the elliptic integral of the second kind 
\begin{equation}
E(\kappa) = \int_{0}^{\pi/2} d\theta \sqrt{1-\kappa^{2}\sin^{2}(\theta)} \;.
\end{equation}
In this work, the elliptic modulus is defined by
\begin{equation}
\kappa = \frac{2\sinh(2\eta)}{\cosh^{2}(2\eta)} \;,
\end{equation}
which satisfies the relation $\kappa^2+(\kappa^{\prime})^2 = 1$, where $\kappa^{\prime}$ is given by
\begin{equation}
\kappa^{\prime} = 2\tanh^2(2\eta)-1 \;,
\end{equation}
and where
\begin{equation}
\eta = \frac{1}{2} \ln\left[\cosh(2\psi)e^{2\phi}\right]  \;,
\label{eq:eta}
\end{equation}
with $\phi = \beta J$ and $\psi = \beta H$. Let $\xi = [\beta,H]$. The derivative of $\kappa$ and $\kappa^{\prime}$ with respect to $\xi$ is
\begin{equation}
\frac{\partial \kappa}{\partial \xi} = -4\kappa^{\prime}\text{sech}(2\eta) \frac{\partial \eta}{\partial \xi}  \;,
\end{equation}
and 
\begin{equation}
\frac{\partial \kappa^{\prime}}{\partial \xi} = 4\kappa\text{sech}(2\eta) \frac{\partial \eta}{\partial \xi}  \;.
\end{equation}
Note further the relation which is useful in the derivation of the system's thermodynamics
\begin{equation}
\frac{1}{\kappa}\frac{\partial \kappa}{\partial \xi} = -2\kappa^{\prime}\text{coth}(2\eta)  \frac{\partial \eta}{\partial \xi} \;.
\end{equation}
Furthermore, the derivative of $\eta$ with respect to $\beta$ and $H$ is 
\begin{equation}
\frac{\partial \eta}{\partial \beta} = J+H \tanh(2\beta H) \;,
\end{equation}
and 
\begin{equation}
\frac{\partial \eta}{\partial H} = \beta \tanh(2\beta H) \;.
\end{equation}

\clearpage
\newpage
\section*{References}
\bibliographystyle{apsrev4-2}
\bibliography{references.bib}

\clearpage
\newpage
\begin{table}[!t]
\renewcommand{\arraystretch}{1.2} 
\renewcommand{\tabcolsep}{0.2cm}
\begin{tabularx}{\textwidth}{rX}
\hline\hline
\multicolumn{2}{l}{Algorithm used in BKL sampling}\\
\hline\hline
- & Assume that the transition probabilities $P_{j}$ are known.\\
- & Calculate rates $Q_{i}=\sum_{j=1}^{i}N_{j}P_{j}$, where $N_{j}$ is the number of spins in class $j$. \\
- & Generate two uniformly distributed random numbers $u_{1}$ $\in$ $[0,1]$ and $u_{2}$ $\in$ $[0,1]$. \\
- & Identify the class $k$ of the transition to be performed $Q_{k-1}$ $\leq$ $u_{1}Q_{max}$ $<$ $Q_{k}$, with $Q_{0} = 0$. \\
- & Identify a spin to update in the chosen class by generating a uniformly distributed random integer $u_{3}$ $\in$ $[0,N_{k}]$. \\
- & Perform the chosen transition and update $N$ in each class.\\
- & Update the time by an amount $\Delta t = -\ln(u_{2})/Q_{max}$. \\
\hline
\end{tabularx}
\begin{tabularx}{\textwidth}{cccXcccX}
\hline\hline
\multicolumn{8}{l}{Transition probabilities in two dimensions} \\ \hline\hline
Class & $S_{i}$ & nn in $+1$ & Transition energy & Class & $S_{i}$ & nn in $+1$ & Transition energy\\ \hline
1 & +1 & 4 & $\Delta E = 8J + 2H$ & 6 & -1 & 4 & $\Delta E = -8J - 2H$\\
2 & +1 & 3 & $\Delta E = 4J + 2H$ & 7 & -1 & 3 & $\Delta E = -4J - 2H$\\
3 & +1 & 2 & $\Delta E = 2H$ & 8 & -1 & 2 & $\Delta E = -2H$\\
4 & +1 & 1 & $\Delta E = -4J + 2H$ & 9 & -1 & 1 & $\Delta E = 4J - 2H$\\
5 & +1 & 0 & $\Delta E = -8J + 2H$ & 10 & -1 & 0 & $\Delta E = 8J - 2H$ \\
\hline
\end{tabularx}
\caption{Algorithm used in BKL sampling and transition probabilities used in the algorithm in two dimensions. The class, the initial state $S_{i}$ of the spin to be updated, the number of nearest neighbors (nn) in the state $+1$ and transition energies are shown.}
\label{tab:transition}
\end{table}

\clearpage
\newpage
\begin{figure}
\includegraphics[width=\textwidth]{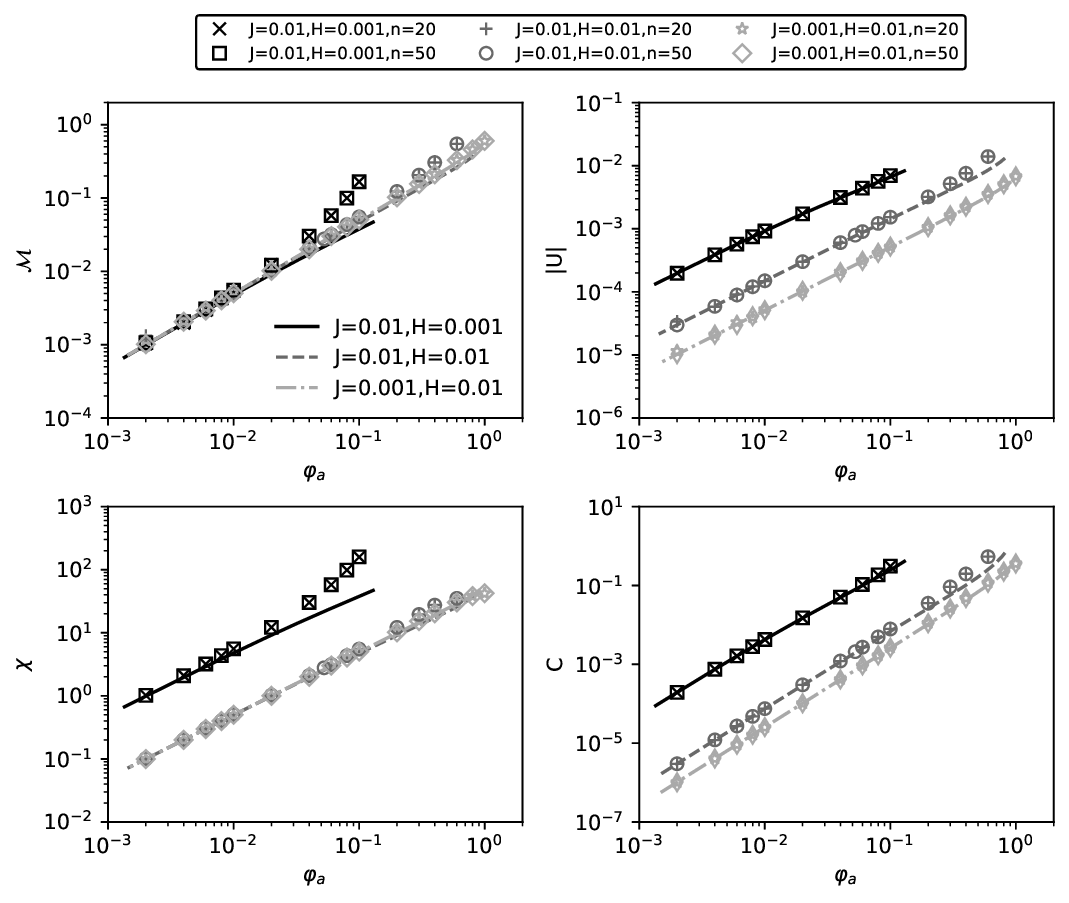}
\caption{Magnetization ($\mathcal{M}$), susceptibility ($\chi$), absolute value of the internal energy ($|U|$), and specific heat ($C$) of the two-dimensional Ising model as a function of the angle $\varphi_{a} = \text{arctan}[\sinh(2\beta H)\exp(2\beta J)]$. Results obtained with the approximate expressions (lines) and numerical simulations of lattice sizes $N = n \times n$  where $n = [20,50]$ (symbols) are compared. The cases where $J = 0.001 < H = 0.01$ (solid lines), $J = H = 0.01$ (dashed  lines), and $J = 0.01 > H = 0.001$ (dash-dotted lines) are shown. The legends apply to all plots.}
\label{fig:varphi_2D}
\end{figure}

\clearpage
\newpage
\begin{figure}
\includegraphics[width=\textwidth]{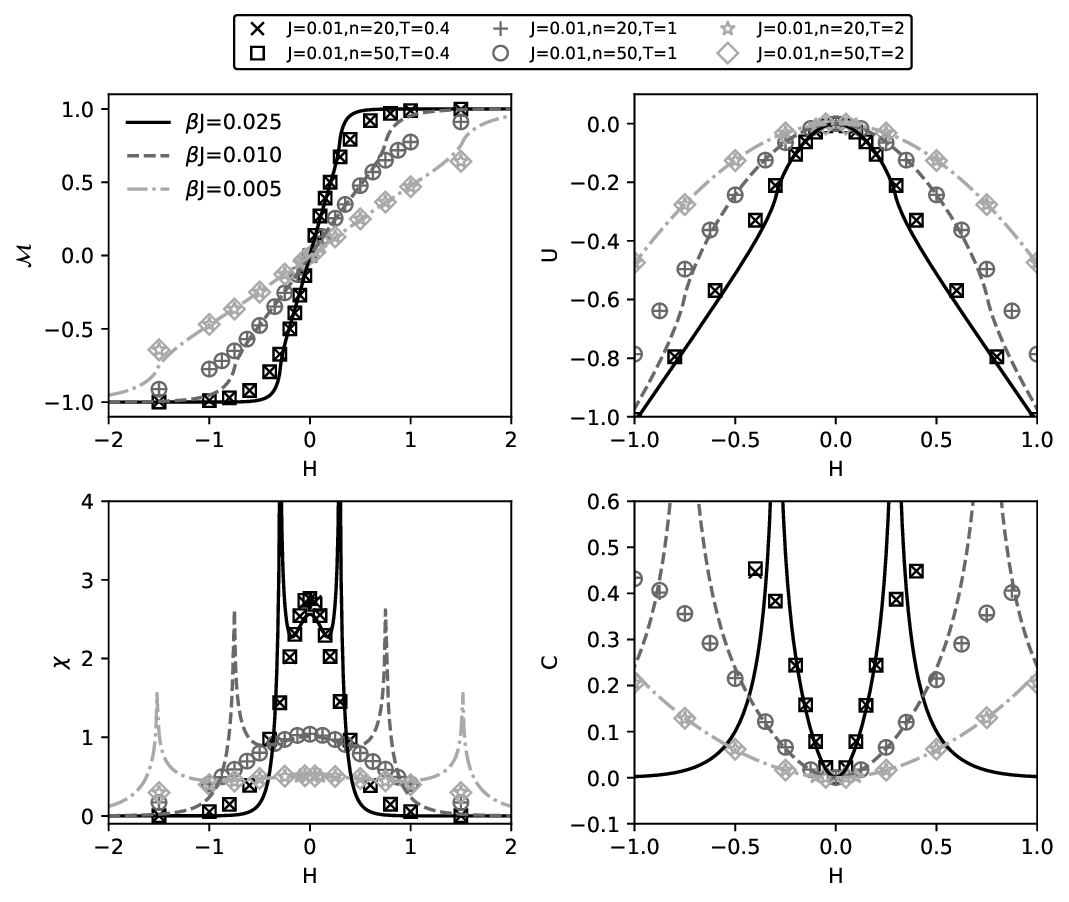}
\caption{Magnetization ($\mathcal{M}$), susceptibility ($\chi$), internal energy ($U$), and specific heat ($C$) of the two-dimensional Ising model shown as a function of the external magnetic field strength $H$. Results obtained with the approximate expressions (solid, dashed, and dash-dotted lines at $T$=0.4 ($\varphi_{a} \approx 0.05$), 1 ($\varphi_{a} \approx 0.02$), and $2$ ($\varphi_{a} \approx 0.01$), respectively) and numerical simulations of lattice sizes $N = n \times n$ where $n = [20,50]$ (symbols) are compared. The interaction energy is fixed at $J = 0.01$. The legends apply to all plots.}
\label{fig:thermo2D_J001}
\end{figure}

\clearpage
\newpage
\begin{figure}
\includegraphics[width=\textwidth]{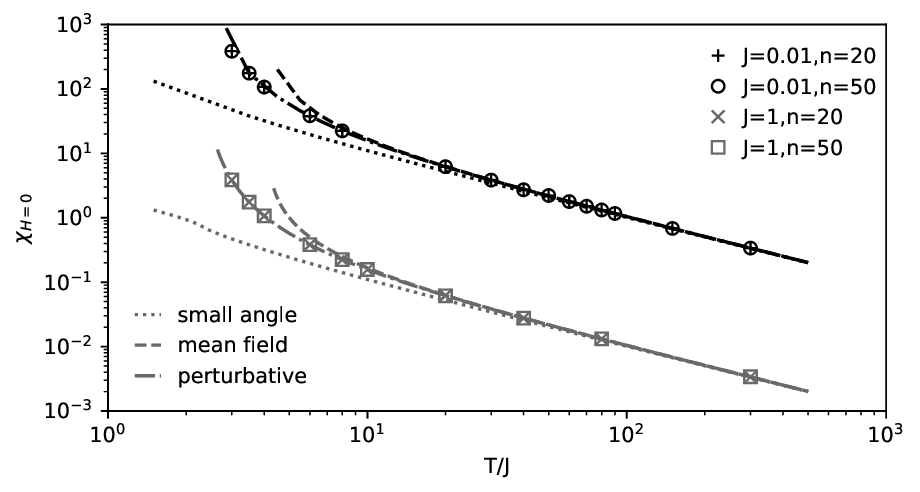}
\caption{Susceptibility ($\chi_{H=0}$)  of the two-dimensional Ising model with $H = 0$ as a function of the reduced temperature $T/J$, where $T$ is the temperature and $J$ is the interaction energy between neighboring spins. Results obtained with the approximate expressions [small angle (dotted lines), mean-field theory (dashed lines), and perturbative expansion (dash-dotted lines)] and numerical simulations of lattice sizes $N = n \times n$ where $n = [20,50]$ (symbols) are shown. The temperature was varied  for both values of $J$ ($J=0.01 $ and $J=1$) considered.}
\label{fig:susceptibility_H0}
\end{figure}

\clearpage
\newpage

\section*{Supplementary material:
Small angle approximation applied to the one-dimensional Ising model in an external magnetic field}

For completeness, we apply the small angle approximation introduced in this work to the one-dimensional Ising model in an external magnetic field. By using its quaternion representation, the model is first exactly solved to obtain  expressions for the physical quantities (magnetization, susceptibility, internal energy, and specific heat). The small angle approximation is then applied, and the results are compared with the exact values and numerical simulations.


\subsection{One-dimensional Ising model with the quaternion algebra} \label{sec:exact_1dIsing}

Consider a linear chain of $n$ spins with state $S = \pm 1$ and periodic boundary conditions $S_{n+1} = S_{1}$. Let $E$ be the interaction energy of the linear chain, defined by
\begin{equation}
E = -J\sum_{i=1}^{n} S_{i}S_{i+1} -\frac{H}{2}\sum_{i=1}^{n} \left(S_{i}+S_{i+1}\right) \;.
\label{eq:1d_interactionEnergy_approx}
\end{equation}
The partition function $Z$ is therefore given by
\begin{equation}
Z =  \sum_{S_{1}}\cdots\sum_{S_{n}} \exp \left( -\beta E \right) \;.
\label{eq:1d_partition}
\end{equation} 
In one dimension, $Z$ can be written as a two-dimensional matrix named the transfer matrix $\mathcal{T}$. In the thermodynamic limit ($n \rightarrow \infty$), $\mathcal{T}$ is dominated by its largest eigenvalue $\lambda_{max}$. Therefore, the thermodynamics of the one-dimensional Ising model are computed by diagonalizing the transfer matrix and considering its largest eigenvalue. Alternatively, the largest eigenvalue may be obtained by using the quaternion representation of Eq. (\ref{eq:1d_partition}). This concept was first introduced by Kaufman in the diagonalization process of the two-dimensional model with $H=0$ \cite{re:kaufman49a}. To do so, we write the transfer matrix $\mathcal{T}$ as
\begin{equation}
\mathcal{T} = \cosh(\beta H)e^{\beta J} \mathds{1} + e^{-\beta J} X+ \sinh(\beta H)e^{\beta J} Z  \;,
\label{eq:1d_transfer2}
\end{equation} 
where 
\begin{equation} 
\mathds{1} = \begin{bmatrix} 1 & 0 \\ 0 & 1 \end{bmatrix}  ,  \quad X = \begin{bmatrix} 0 & 1 \\ 1 & 0 \end{bmatrix} \;, Y = \begin{bmatrix} 0 & -i \\ i & 0 \end{bmatrix}  ,  \quad Z = \begin{bmatrix} 1 & 0 \\ 0 & -1 \end{bmatrix} \;,
\label{eq:pauli_1d}
\end{equation} 
are the Pauli matrices and are used as basis in the quaternion representation. Factorize the term proportional to $\mathds{1}$ and write Eq. (\ref{eq:1d_transfer2}) as,
\begin{equation}
\mathcal{T}_{\pm} = \cosh(\beta H)e^{\beta J} \left[ \mathds{1} \pm \frac{\text{sec}(\varphi)}{\cosh(\beta H)e^{2\beta J}} X^{\prime}  \right]\;,
\end{equation}
where we have defined, 
\begin{equation}
X^{\prime} =  \cos(\varphi) X + \sin(\varphi) Z \;,
\end{equation}
and
\begin{align}
\cos(\varphi) &= \frac{1}{\sqrt{1+\sinh^{2}(\beta H)e^{4\beta J}}} \;,  \\
\sin(\varphi) &=  \frac{\sinh(\beta H)e^{2\beta J}}{\sqrt{1+\sinh^{2}(\beta H)e^{4\beta J}}} \;,
\end{align} 
and $\tan(\varphi)$ = $\sinh(\beta H)e^{2\beta J}$. Here, $\mathcal{T}_{\pm}$ represents the two equivalent solutions corresponding to the two choices of sign in front of $X^{\prime}$; because $(X^{\prime})^2 = \mathds{1}$, both choices lead to the same eigenvalues. In these terms, the angle $\varphi$ represents a rotation between the reference frame $(X,Z)$ and a rotated frame $(X^{\prime},Z^{\prime})$ so that
\begin{equation}
\begin{bmatrix}  X^{\prime} \\ Z^{\prime} \end{bmatrix} = \begin{bmatrix} \cos(\varphi) & \sin(\varphi)  \\ -\sin(\varphi) & \cos(\varphi)  \end{bmatrix} \begin{bmatrix}  X \\ Z \end{bmatrix} \;.
\end{equation} 
The transfer matrix is then given by
\begin{equation}
\mathcal{T}_{\pm} = \cosh(\beta H)e^{\beta J} \left[ \mathds{1} \pm \tanh(\omega) X^{\prime} \right] \;,
\label{eq:1d_transfer7}
\end{equation} 
where we have defined
\begin{equation}
\tanh(\omega) = \frac{\text{sec}(\varphi)}{\cosh(\beta H)e^{2\beta J}} \;. 
\label{eq:tanh_omega}
\end{equation}
Since $\left[\cos(\varphi) X + \sin(\varphi) Z\right]^{2} = 1$ and using the exponential matrix definition $e^{b A} = \cosh(b)\mathds{1} +\sinh(b)A$, where $A$ is matrix satisfying $A^2 = 1$, and where $b$ is a constant,   Eq. (\ref{eq:1d_transfer7}) can be written as 
\begin{equation}
\mathcal{T}_{\pm} = \frac{\cosh(\beta H)e^{\beta J} }{\cosh(\omega)} e^{\pm\omega X^{\prime}} \;.
\label{eq:1d_transfer6}
\end{equation}
Introduce two matrices $\Gamma_{1} = Z^{\prime}$ and $\Gamma_{2} = Y$ that follow the anticommutation rule $\Gamma_{p}\Gamma_{q}+ \Gamma_{q}\Gamma_{p} = 2\delta_{pq}$, where $\delta_{pq} = 1$  if $p=q$ and $\delta_{pq} = 0$ otherwise. In these terms, Eq. (\ref{eq:1d_transfer6}) is 
\begin{equation}
\mathcal{T}_{\pm} = \frac{\cosh(\beta H)e^{\beta J} }{\cosh(\omega)} e^{\mp i\omega \Gamma_{2}\Gamma_{1}} \;.
\label{eq:1d_transfer4}
\end{equation}
The quantity $\exp{\left(\mp i\omega \Gamma_{2}\Gamma_{1}\right)}$ is the spin representative of the rotation $\omega$ with eigenvalue $\exp(\pm \omega)$ \cite{re:kaufman49a}. The largest eigenvalue $\lambda_{max}$ of both $\mathcal{T}_{+}$ and $\mathcal{T}_{-}$ is 
\begin{equation}
\lambda_{max} = \frac{\cosh(\beta H)e^{\beta J} }{\cosh(\omega)} e^{\omega} \;,
\end{equation}
and therefore, the  free energy $(\beta F)$ is given by
\begin{equation}
(\beta F) = \beta J +\ln[\cos(\varphi)]- \ln\left[ \text{coth}(\omega)+1\right]\;.
\label{eq:free_1d}
\end{equation}
Thermodynamic quantities can be obtained from the free energy. The magnetization per spin $\mathcal{M}$ is given by \cite{re:mccoy73}
\begin{equation}
\mathcal{M} =  -\frac{\partial F}{\partial H}  = \sin(\varphi)\;.
\label{eq:m_1d}
\end{equation}
Furthermore, the susceptibility per spin $\chi$ is 
\begin{equation}
\chi =  \frac{\partial \mathcal{M}}{\partial H}  = \beta \cos^2(\varphi) \text{coth}(\omega) \;.
\label{eq:chi_1d}
\end{equation}
Moreover, the internal energy per spin $U$ is given by
\begin{equation}
U = \frac{\partial (\beta F)}{\partial \beta} = J  \left[\frac{\cos(2\varphi)-\text{coth}(\omega)}{1+\text{coth}(\omega)} \right]  - H \mathcal{M} \;.
\label{eq:u_1d}
\end{equation}
Finally, the specific heat per spin is
\begin{equation}
\begin{split}
C = -\beta^{2}\frac{\partial U }{\partial \beta} &= 4\beta^2 \cos^{2}(\varphi) \coth(\omega) \times \\
& \quad\left\{ J^2\frac{\left(1+\sin^{2}(\varphi)\left[1+2\tanh(\omega)\right] \right)}{\left[1+\coth(\omega)\right]^2}  +   JH \frac{\sin(\varphi)}{\coth(\omega)} +  \frac{H^2}{4} \right\} \;.
\end{split}
\label{eq:c_1d}
\end{equation}
The thermodynamic quantities [Eqs. (\ref{eq:m_1d}) - (\ref{eq:c_1d})] are shown in Figs. (\ref{fig:varphi_1D}) - (\ref{fig:thermo1D_varyT}).

\subsection{Approximate thermodynamics of the one-dimensional Ising model}

In this section, we introduce the small angle approximation and apply it to the one-dimensional model. We rescale $H \to 2H$ to account for the two solutions of Eq. (\ref{eq:1d_transfer6}). In these terms, the small angle approximation is defined as
\begin{equation}
\varphi \to 0 \Rightarrow  \varphi_{a} = \lim_{\begin{subarray}{l}
\beta H \to 0 \\
\beta J \to 0
\end{subarray}} \text{arctan}\left[ \sinh(2\beta H)e^{2\beta J} \right] \;,
\end{equation}
where $a$ stands for approximate. Therefore, Eq. (\ref{eq:tanh_omega}) becomes
\begin{equation}
\text{coth}(\omega_{a}) = \cosh(2\beta H)e^{2\beta J} \;,
\label{eq:tanhOmega_approx}
\end{equation}
In these terms, the approximate free energy $(\beta F)_{a}$ is 
\begin{equation}
(\beta F)_{a} = \beta J - \ln\left[ \text{coth}(\omega_{a})+1\right]\;. 
\label{eq:freeEnergy_1D_approx}
\end{equation}
The magnetization $\mathcal{M}^{1D}_{a}$ is obtained by taking the derivative of Eq. (\ref{eq:freeEnergy_1D_approx}) with respect to the external field and correcting the results with a factor of 1/2, which leads to 
\begin{equation}
\mathcal{M}^{1D}_{a} = \frac{\sinh(2\beta H)e^{2\beta J}}{1+\cosh(2\beta H)e^{2\beta J}}\;.
\label{eq:ma_1d}
\end{equation}
Furthermore, the approximate magnetic susceptibility ($\chi^{1D}_{a}$) is obtained by taking the derivative of $\mathcal{M}_{a}$  with respect to $H$
\begin{equation}
\chi^{1D}_{a} =  2\beta e^{4\beta J}\frac{\left[1+\cosh(2\beta H)e^{-2\beta J}\right]}{\left[1+\cosh(2\beta H)e^{2\beta J}\right]^{2}}\;.
\label{eq:chia_1d}
\end{equation}
Similarly to the magnetization, the approximate internal energy per spin $U^{1D}_{a}$ is obtained by taking the derivative of the free energy with respect to $\beta$ and correcting the magnetization by a factor of 1/2, which leads to
\begin{equation}
U^{1D}_{a} = J\left[\frac{1-\cosh(2\beta H)e^{2\beta J}}{1+\cosh(2\beta H)e^{2\beta J}}\right] -H\mathcal{M}_{a}\;.
\label{eq:ua_1d}
\end{equation}
Finally, the approximate specific heat per spin ($C^{1D}_{a}$) is
\begin{equation}
C^{1D}_{a} =  2\beta^{2}e^{2\beta J} \left\{ \frac{ 2J^2\cosh(2\beta H)+3JH\sinh(2\beta H) + H^2\left[ \cosh(2\beta H)+ e^{2\beta J}\right] }{\left[1+\cosh(2\beta H)e^{2\beta J}\right]^2} \right\}  \;. 
\label{eq:ca_1d}
\end{equation}
Interestingly as $J = 0$, the approximate expressions [Eqs. (\ref{eq:ma_1d}) - (\ref{eq:ca_1d})] are identical to the exact expressions [Eqs. (\ref{eq:m_1d}) - (\ref{eq:c_1d})]. 

\subsection{Numerical simulations} 

The one-dimensional Ising model was simulated by using the BKL algorithm as described in the manuscript. Let $\mathcal{S}$  be the sum of all spin states of a configuration normalized by the total number of spins computed as 
\begin{equation}
\mathcal{S} = \frac{1}{n}\sum_{i=1}^{n}S_{i}  \;.
\end{equation}		
The magnetization $\mathcal{M}$ and the magnetic susceptibility $\chi$ were computed by calculating the average and variance of $\mathcal{S}$ over all configurations such that
$\mathcal{M} = \langle \mathcal{S} \rangle$ and $\chi = \beta\langle \Delta \mathcal{S}^2 \rangle$. Furthermore, let $\mathcal{E}$ be the energy per spin of a configuration computed as 
\begin{equation}
\mathcal{E} = -\frac{J}{2n}\sum_{i=1}^{n}\left(S_{i}S_{i-1}+S_{i}S_{i+1}\right) -\frac{H}{n}\sum_{i=1}^{n}S_{i}  \;, 
\end{equation}
where a factor of 1/2 was applied to account for double counting. The internal energy $U$ and the specific heat $C$ were computed with $U = \langle \mathcal{E} \rangle$ and $C = \beta^{2}\langle \Delta \mathcal{E}^2 \rangle$. The simulations were performed for a chain length of $N=\{500, 1000\}$ with $\Delta t_{therma}$ = 5 $\times$ 10$^{5}$ and $\Delta t_{compute}$ = 10$^{3}$. A total of 96,000 configurations were simulated.

\subsection{Comparison between analytical and numerical results}

\begin{figure}
\includegraphics[width=\textwidth]{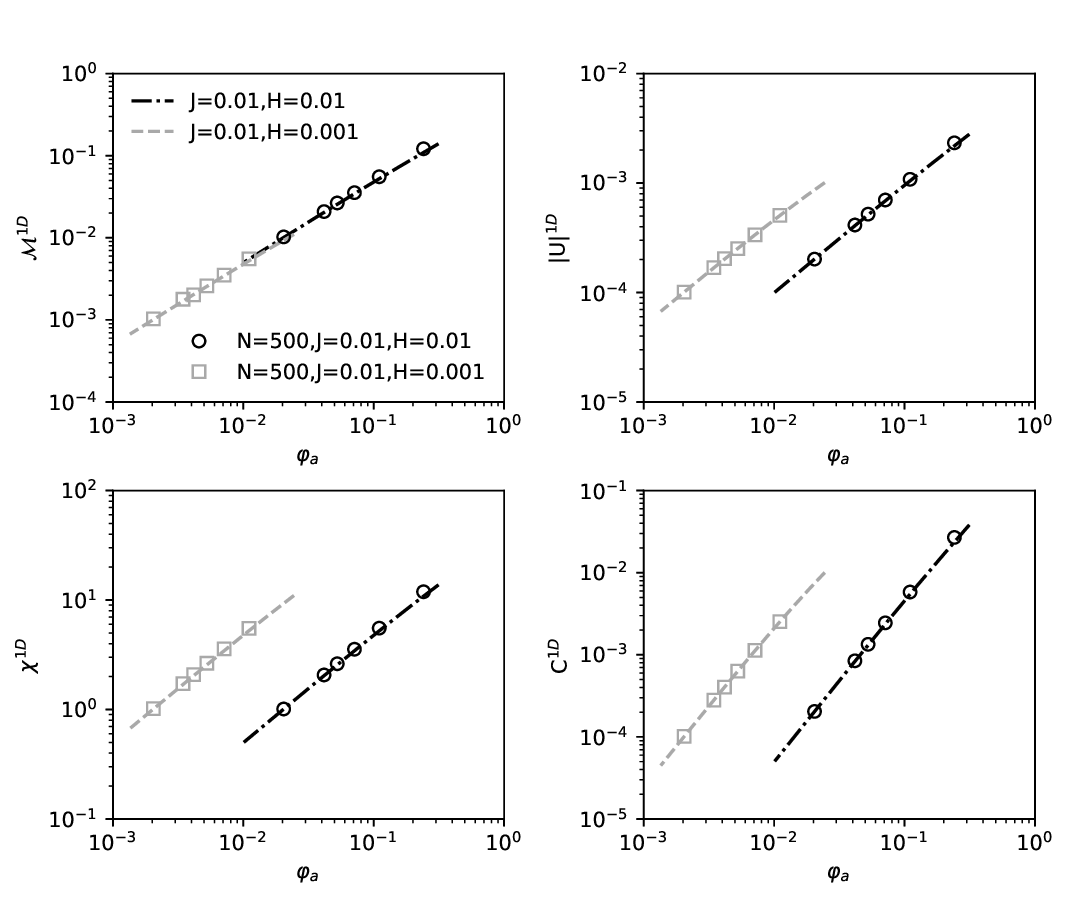}
\caption{Magnetization ($\mathcal{M}^{1D}$), susceptibility ($\chi^{1D}$), absolute value of the internal energy ($|U|^{1D}$), and specific heat ($C^{1D}$) of the one-dimensional Ising model as a function of the angle $\varphi_{a} = \text{arctan}[\sinh(2\beta H)\exp(2\beta J)]$. Results obtained with the approximate expressions (lines) and numerical simulations of a linear chain of size $N = 500$ (symbols) are compared. The cases where $J = H = 0.01$ (dash-dotted lines), and $J = 0.01 > H = 0.001$ (dashed  lines) are shown. The legends shown apply to all plots.}
\label{fig:varphi_1D}
\end{figure}

The approximate thermodynamic expressions [Eqs. (\ref{eq:ma_1d}) - (\ref{eq:ca_1d})] are compared with numerical simulations as the angle $\varphi_{a}$ is varied in Fig. \ref{fig:varphi_1D}. The cases where  $J = H$  and $J > H$ were considered. The agreement between the approximate expressions and the numerical values was excellent in regions where the approximation was valid.

\begin{figure}[!t]
\includegraphics[width=6in]{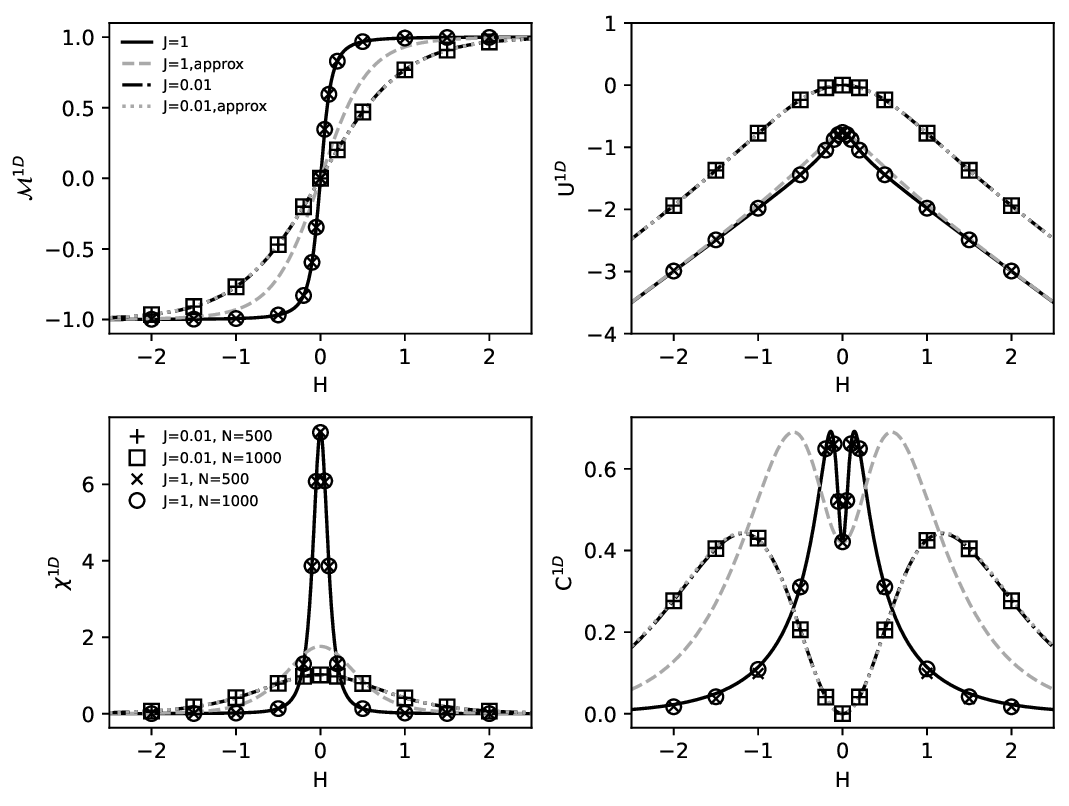}
\caption{Magnetization ($\mathcal{M}^{1D}$), susceptibility ($\chi^{1D}$), internal energy ($U^{1D}$), and specific heat ($C^{1D}$) of the one-dimensional Ising model obtained with the exact and approximate expressions (lines) and the numerical simulations of a linear chain of size $N = [500,1000]$ (symbols) as the external field strength $H$ is varied. The temperature is fixed at T = 1. The legends shown apply to all plots.}
\label{fig:thermo1D_varyJ}
\end{figure}

The approximate expressions for the thermodynamics of the linear chain as the temperature is kept fixed ($T=1$) for two values of the interaction energy $(J = [0.01,1])$ are shown in Fig. \ref{fig:thermo1D_varyJ} and compared to the exact values and numerical simulations. With $J = 0.01$, the mean absolute difference between approximate and exact values was lower than $5 \times 10^{-5}$ in $H = [-2.5,2.5]$ for all thermodynamic expressions. The approximate and exact values did not agree with each other with $J = 1$ (except for the internal energy), which was expected as the small angle approximation is not valid at large $\phi$. 

\begin{figure}[!t]
\includegraphics[width=6in]{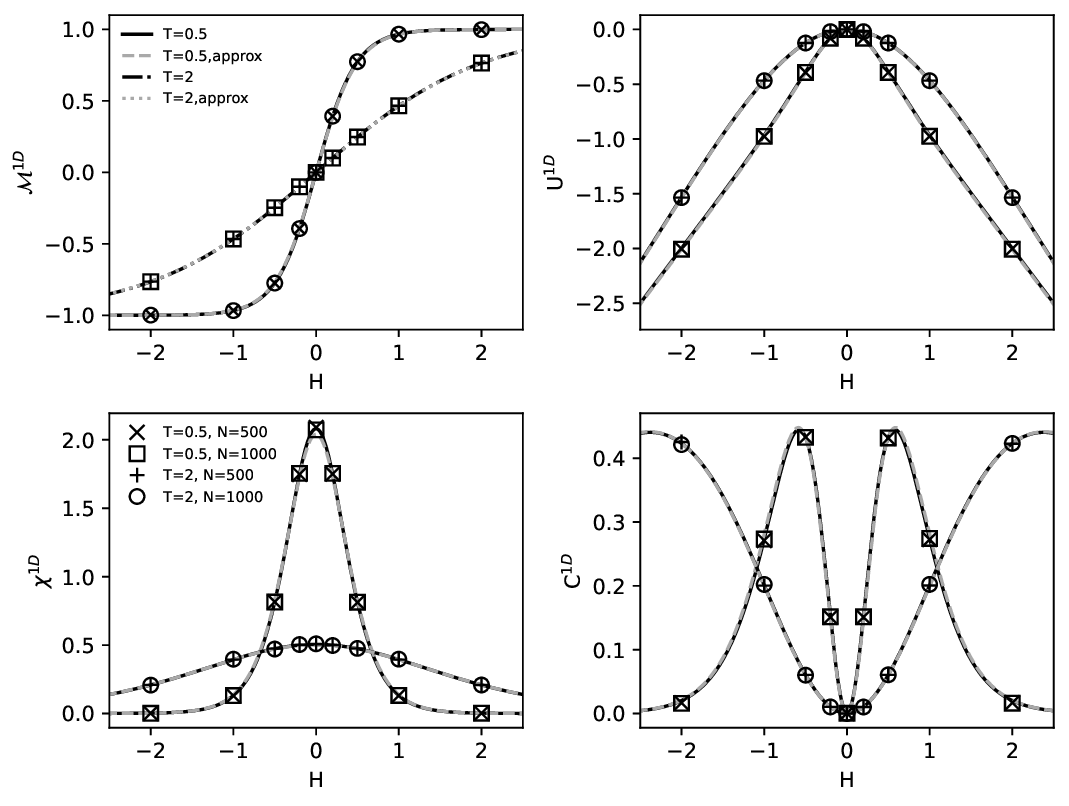}
\caption{Magnetization ($\mathcal{M}^{1D}$), susceptibility ($\chi^{1D}$), internal energy ($U^{1D}$), and specific heat ($C^{1D}$) of the one-dimensional Ising model obtained with exact and approximate expressions (lines) and numerical simulations of a linear chain of size $N = [500,1000]$ (symbols) as the external field $H$ is varied. The interaction energy is fixed at J = 0.01. The legends shown apply to all plots.}
\label{fig:thermo1D_varyT}
\end{figure}

The approximate expressions are shown in Fig. \ref{fig:thermo1D_varyT} as the interaction energy is fixed ($J = 0.01$) for two values of the temperature ($T = [0.5,2]$) and compared to the exact values and numerical simulations. In this case, the mean absolute difference between approximate and exact values was lower than $3 \times 10^{-3}$ for all thermodynamic quantities in $H = [-2.5,2.5]$ and the two temperatures considered. It is worth noting that in Fig. \ref{fig:thermo1D_varyT}, $J < |H|$ in almost all points shown.

\newpage
\section{Validation of the numerical results in two-dimensions} \label{sec:validation}

\begin{figure}[!t]
\includegraphics[width=6in]{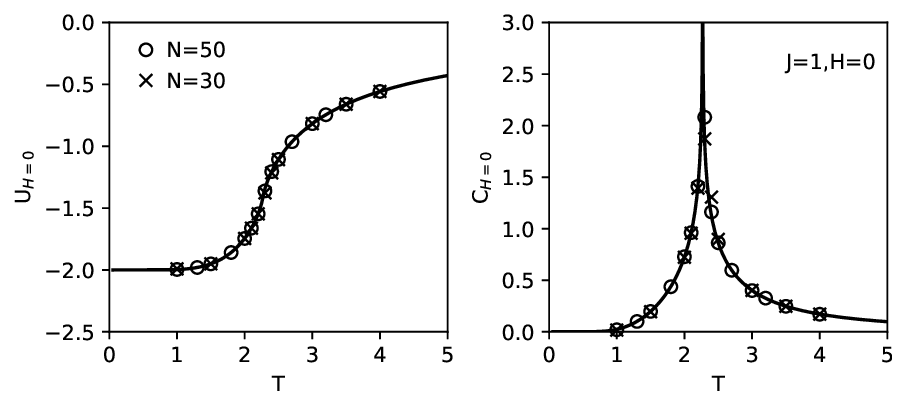}
\caption{Internal energy ($U_{H=0}$) and specific heat ($C_{H=0}$) of the two-dimensional Ising model at $H = 0$ obtained with the exact expressions (solid line) and numerical simulations of lattice sizes $N = n \times n$ where $n = [30,50]$ (symbols). The interaction energy is fixed at J = 1. The legends shown apply to both plots.}
\label{fig:thermo2D_H0}
\end{figure}

\begin{figure}[!t]
\includegraphics[width=6in]{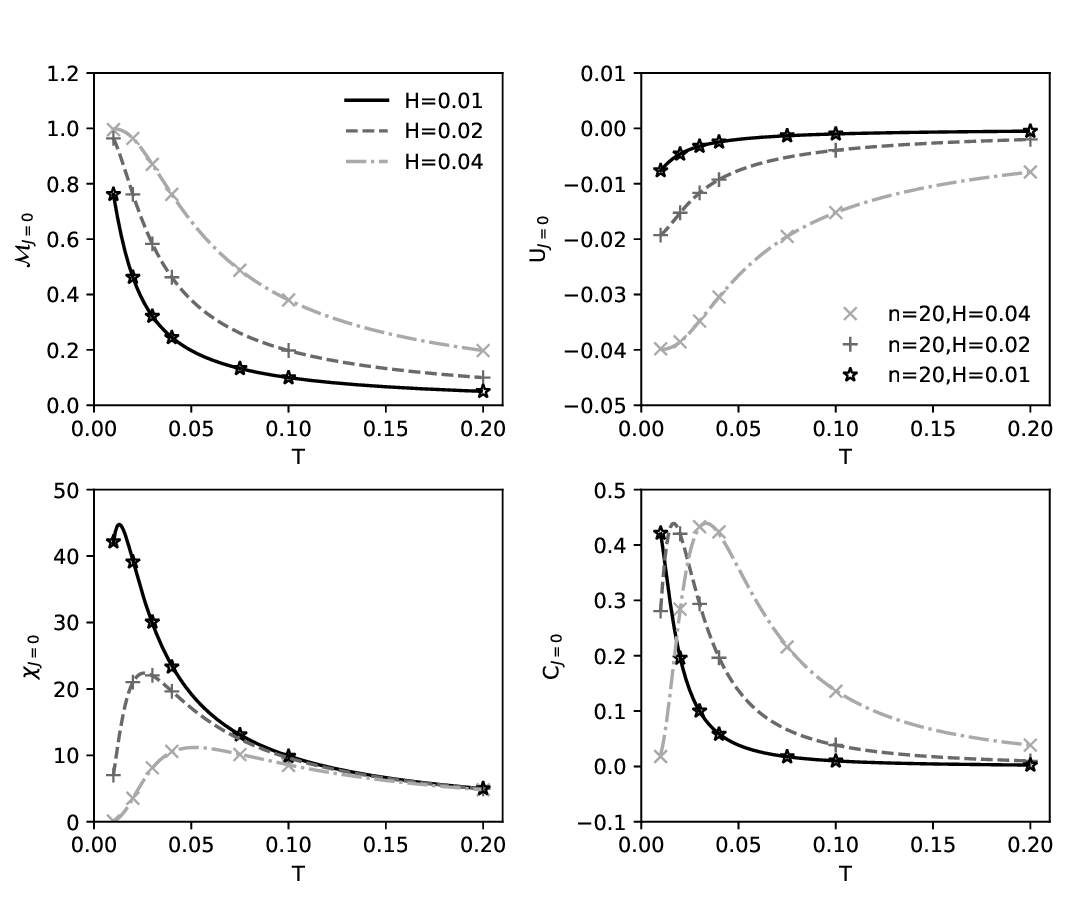}
\caption{
Magnetization ($\mathcal{M}_{J=0}$), susceptibility ($\chi_{J=0}$), internal energy ($U_{J=0}$), and specific heat ($C_{J=0}$) of the two-dimensional Ising model at $J = 0$ obtained with the exact expressions (solid and dashed lines) and numerical simulations of lattice sizes $N = n \times n$ where $n = 20$ (symbols) as the temperature $T$ is varied. The legends shown apply to all plots.}
\label{fig:thermo2D_J0}
\end{figure}

To validate the simulations, numerical results for the cases where $H=0$ and $J=0$ are compared to the exact expressions as the temperature is varied in Fig. \ref{fig:thermo2D_H0} and \ref{fig:thermo2D_J0}, respectively. At $H = 0$, the internal energy is
\begin{equation}
U_{H=0} = -J \text{coth}(2\beta J)\left[1+\frac{2}{\pi}\kappa^{\prime}K(\kappa)\right]\;,
\label{eq:ua_2d_exact}
\end{equation}
where 
\begin{equation}
\kappa^{\prime} = 2\tanh^{2}(2\beta J)-1 \;,
\end{equation}	
where $K$ is the complete elliptic integral of the first kind
\begin{equation}
K(\kappa) = \displaystyle{\int_{0}^{\pi/2}\frac{d\theta}{\sqrt{1-\kappa^{2}\sin^{2}(\theta)}}} \;,
\end{equation}
and where 
\begin{equation}
\kappa = \frac{2\sinh(2\beta J)}{\cosh^{2}(2\beta J)} \;. 
\end{equation}
The specific heat is given by
\begin{equation}
C = \frac{2}{\pi} (\beta J)^{2}\text{coth}^{2}(2\beta J)\bigg\{ 2[K(\kappa)-E(\kappa)]-(1-\kappa^{\prime})\left[\frac{\pi}{2}+\kappa^{\prime}K(\kappa) \right] \bigg\} \;.
\label{eq:ca_2d_exact}
\end{equation}
At $J = 0$, the two-dimensional model behaves as an ideal paramagnet for which the magnetization, susceptibility, internal energy, and specific heat are respectively given by
\begin{equation}
\mathcal{M}_{J=0} = \tanh(\beta H)\;,
\end{equation}
\begin{equation}
\chi_{J=0} = \beta\text{sech}^{2}(\beta H)\;,
\end{equation}
\begin{equation}
U_{J=0} = -H \tanh(\beta H)\;,
\end{equation}
and 
\begin{equation}
C_{J=0} = (\beta H)^2\text{sech}^{2}(\beta H)\;.
\end{equation}




\clearpage
\newpage
\begin{table}[!t]
\renewcommand{\arraystretch}{1.2} 
\renewcommand{\tabcolsep}{0.2cm}
\begin{tabularx}{\textwidth}{rX}
\hline\hline
\multicolumn{2}{l}{Algorithm used in BKL sampling}\\
\hline\hline
- & Assume that the transition probabilities $P_{j}$ are known\\
- & Calculate rates $Q_{i}=\sum_{j=1}^{i}N_{j}P_{j}$, where $N_{j}$ is the number of spins in class $j$ \\
- & Generate two uniformly distributed random numbers $u_{1}$ $\in$ $[0,1]$ and $u_{2}$ $\in$ $[0,1]$ \\
- & Identify the class $k$ of the transition to be performed: $Q_{k-1}$ $\leq$ $u_{1}Q_{max}$ $<$ $Q_{k}$, with $Q_{0} = 0$ \\
- & Identify a spin to update in the chosen class by generating a uniformly distributed random integer $u_{3}$ $\in$ $[0,N_{k}]$ \\
- & Perform the chosen transition and update $N$ in each class\\
- & Update the time by an amount $\Delta t = -\ln(u_{2})/Q_{max}$ \\
\end{tabularx}
\begin{tabularx}{\textwidth}{cccXcccX}
\hline\hline
\multicolumn{8}{l}{Transition probabilities in one dimension} \\ \hline\hline
Class & $S_{i}$ & nn in $+1$ & Transition energy & Class & $S_{i}$ & nn in $+1$ & Transition energy\\ \hline
1 & +1 & 2 & $\Delta E = 4J + 2H$ 	& 4 & -1 & 2 & $\Delta E = -4J - 2H$\\
2 & +1 & 1 & $\Delta E = 2H$ 				& 5 & -1 & 1 & $\Delta E = - 2H$\\
3 & +1 & 0 & $\Delta E = -4J + 2H$ 	& 6 & -1 & 0 & $\Delta E = 4J - 2H$\\
\hline
\end{tabularx}
\caption{Algorithm used in BKL sampling and transition probabilities used in the algorithm in one dimension. The class, the initial state $S_{i}$ of the spin to be updated, the number of nearest neighbors (nn) in the state $+1$ and transition energies are shown.}
\label{tab:transition_1D}
\end{table}


\clearpage
\newpage

\clearpage
\newpage

\clearpage
\newpage

\clearpage
\newpage

\clearpage
\newpage


\end{document}